\documentclass[a4paper]{article} 
\usepackage[top=2cm, bottom=2cm, left=1.9cm, right=1.9cm]{geometry} 
\usepackage[super,square,comma,sort&compress]{natbib}
\usepackage{textcomp,gensymb}
\usepackage{siunitx}
\newcommand{\calciumforty}{$^{40}\mathrm{Ca}^+$}
\usepackage{comment}
\usepackage{authblk}
\usepackage{graphicx}
\usepackage[hidelinks]{hyperref}
\usepackage{color,soul}
\renewcommand\hl[1]{#1} 

\author[1,2]{Matthias Dietl\thanks{Corresponding author; Electronic address: \url{Matthias.Dietl@infineon.com}}}
\author[2]{Marco Valentini}
\author[1,2]{Fabian Anmasser}
\author[1,3]{Alexander Zesar}
\author[1]{Silke Auchter}
\author[4]{Martin van Mourik}
\author[2,4]{Thomas Monz}
\author[2,5]{Rainer Blatt}
\author[1]{Clemens Rössler}
\author[2]{Philipp Schindler}
\affil[1]{Infineon Technologies Austria AG, Siemensstraße 2, A-9500, Villach, Austria}
\affil[2]{Institut für Experimentalphysik, Universität Innsbruck, Technikerstraße 25, A-6020, Innsbruck, Austria}
\affil[3]{Institut für Physik, Universität Graz, Universitätsplatz 5, A-8010 Graz, Austria}
\affil[4]{Alpine Quantum Technologies GmbH, Technikerstraße 17/1, 6020 Innsbruck, Austria}
\affil[5]{Institut für Quantenoptik und Quanteninformation, Österreichische Akademie der Wissenschaften, Technikerstraße 21a, A-6020, Innsbruck, Austria}
\date{}

\begin{document}

\title{\textbf{Test and characterization of multilayer ion traps on fused silica}}
\maketitle

\begin{abstract}
\noindent Ion traps are a promising architecture to host a future quantum computer. Several challenges, such as signal-routing, power dissipation, and fabrication quality, need to be overcome to scale ion trap devices to hundreds of ions. Currently, ion traps are often fabricated on silicon substrates which result in high power dissipation. Substrates that lead to lower power dissipation are preferred. In this work, we present a multi-metal layer ion trap on a fused silica substrate that is fabricated and tested in an industrial facility. Its design and material-stack are tailored to minimize power dissipation. Furthermore, we characterize the integrated temperature sensors and verify functionality down to 10 K. Moreover, we demonstrate an automated wafer test to validate each trap chip prior to its integration into experimental setups. Subsequently, we characterize electric field noise and electric stray fields using a single trapped-ion as a probe, showing an improvement in trap performance over similar trap designs realized on silicon substrates.

\end{abstract}

\section{Introduction}
Ion trap based quantum information processors are among the most advanced platforms to implement a universal quantum computer.\cite{Bruzewicz2019} Despite achieving the highest performance among all platforms in terms of single- and two-qubit gate fidelities,\cite{smith2024singlequbitgateserrors107,löschnauer2024scalablehighfidelityallelectroniccontrol} the ion trap processors demonstrated so far routinely operate with up to about 50 qubits, compared to the hundreds controlled in neutral atom experiments \cite{Ebadi2021,Bluvstein2023} or superconducting systems.\cite{Kim2023,Acharya2024} One of the outstanding challenges for trapped-ion based quantum computers and simulators is to increase the number of trapped ions while maintaining the capability to control hundreds of qubits with high fidelity.   

To scale up the number of ions in an ion trap, multiple trap architectures have been explored. These include the Quantum-Charged-Coupled-Device (QCCD) architecture,\cite{Kielpinski2002, Pino2021, Delaney2024} where the ion qubits are connected via a combination of ion shuttling, ion crystal rotations, and splitting and merging operations. Other approaches are the two-dimensional (2D) lattice architecture,\cite{Tanaka2014, Sterling2014, Bruzewicz2016, Holz2020, valentini2024}, where 2D connectivity is achieved by combining 2D ion transport with long-range Coulomb interactions, and the Penning micro-trap array,\cite{Jain2024} where ion qubits can be arbitrarily transported and connected along three dimensions above the trap surface. All three approaches utilize micro-fabricated surface or 3D ion traps. The local control of ions benefits from so-called island-like electrodes, where the electrodes' routing does not run on the trap surface, requiring the traps to be fabricated on multiple metal layers. Realizing a large scale ion trap quantum computer presents numerous technological challenges. Here, we will address two of them: The power dissipated by the radio frequency (RF) drive required to operate the trap and reliable fabrication of multi-metal layer ion trap chips. 

The chip substrate plays an important role in determining the amount of RF power that is dissipated within the trap.
Silicon is widely used as a substrate for the fabrication of ion traps, which can be attributed to the extensive knowledge and expertise in semiconductor processing on silicon.\cite{Kilby1976, Heywang2004} Established CMOS-compatible processes on silicon substrate facilitate the integration of waveguides,\cite{Mehta2020, Niffenegger2020, Kwon2024} photodetectors \cite{Setzer2021, Todaro2021, Reens2022} or slots for backside loading of ions \cite{Jung2021, Revelle2020, chung2024} within the trap chip. However, silicon has the drawbacks of high RF losses\cite{Yang2006,Krupka2015,Leibrandt2009} and a small bandgap of \SI{1.12}{e\volt},\cite{Kasap2017} which leads to the creation of free charge carriers due to stray-light.\cite{Romaszko2020, Mehta2014} The RF losses in silicon substrates lead to increased heating of both the substrate and the ion trap. Subsequently, a higher trap temperature leads to higher ion heating rates in cryogenic setups.\cite{Bruzewicz2015,Chiaverini2014} The small bandgap of silicon causes lasers to generate free charge carriers that can create time-dependent fields, which may further increase ion heating rates or lead to uncontrolled variations of the trapping potential.\cite{berkeland1998, Harlander2010} Both of these problems can be mitigated by adding a metallic layer between the trap electrodes and the silicon substrate.\cite{Blain2021,Doret2012,Niedermayr2014,Leibrandt2009}. Although this approach reduces RF losses in the substrate and mitigates the generation of free charge carriers, it also leads to an increased RF to ground capacitance. The higher RF capacitance enhances other power dissipation mechanisms: Ohmic losses within the RF electrodes and dielectric losses in the oxide layer located between the RF electrodes and other metal structures.\cite{Hughes2011,Sterk2024,Blain2021} These losses are considered less critical compared to those in silicon or to the generation of free charge carriers in the substrate,\cite{Yang2006,Krupka2015,Leibrandt2009} because they primarily contribute to manageable localized heating, whereas RF losses in silicon lead to significant substrate heating, and the generation of free charge carriers destabilizes the trapping potential. Nevertheless, Ohmic and dielectric losses in the oxide still represent a concern for the realization of a large-scale ion trap quantum computer operated in a cryogenic environment. 

An alternative to silicon substrates are dielectric substrates, such as fused silica, which exhibit low RF losses ($\tan\delta=0.5\times10^{-4}$\cite{Bussey1964}) and large bandgaps of 7.5-\SI{9.6}{e\volt}.\cite{Tan2003} Due to these characteristics, dielectric substrates do not require a shielding layer, effectively reducing trap capacitance. 
Although single and multi-metal layer ion traps fabricated on glass have been successfully manufactured in non-industrial clean rooms,\cite{Amini2010, Kumph2015, BautistaSalvador2019, xu2023,Simeth2023} large-scale industrial production remains a challenge. The tools and equipment used for the processing of silicon substrates are generally not compatible with the requirements of transparent glass wafer processing.\cite{Haley2024} In particular, the different thermal conductivity and thermal expansion coefficients of glass substrates, compared to those of silicon, can lead to fabrication deviations and bow issues during fabrication.\cite{Deng2020,Damon1973,Becker1982,Glassbrenner1964} For these reasons, glass substrates require the development of new wafer-processing techniques.

Another crucial aspect of ion-trap production is reliability, particularly in the fabrication and testing processes. In this context, methods for testing ion trap chips prior to the dicing process and their integration into experimental setups are valuable. 
A typical ion trap testing procedure relies on optical inspection, but with this method, the evaluation of multilayer traps is limited to the uppermost metal layer. For semiconductor fabrication instead, electrical testing of individual chips at the wafer level is a well-established technique for post-fabrication evaluation.\cite{Das2001,Mann2008,Kang2015} We thus opted for this approach and developed an electrical ion trap testing procedure to both identify fabrication defects and evaluate ion traps prior to their packaging. 

In this paper, we present the design, fabrication and characterization of a three-metal layer ion trap realized on a fused silica substrate and fabricated in the industrial facilities of Infineon Technologies in Villach, Austria.
Details about trap design, fabrication, and power dissipation are given in Section \ref{fabrication_design}. 
In Section \ref{wafertest}, we present a method for electrical testing of ion traps at the wafer level, prior to their integration into experimental setups.
In Section \ref{characterization}, we present the characterization of our fused silica trap within an experimental setup. This includes testing the functionality of the integrated temperature sensors, as well as evaluating the trap’s performance with heating rate measurements and stray field analysis. We conclude our work with an outlook on ion traps with low dissipation and possible improvements for wafer level testing in Section \ref{chap:conclucsion}.

\section{Ion trap}\label{fabrication_design}
\subsection{Design and fabrication}
\begin{figure}[t]
    \centering
    \includegraphics[width=0.8\textwidth]{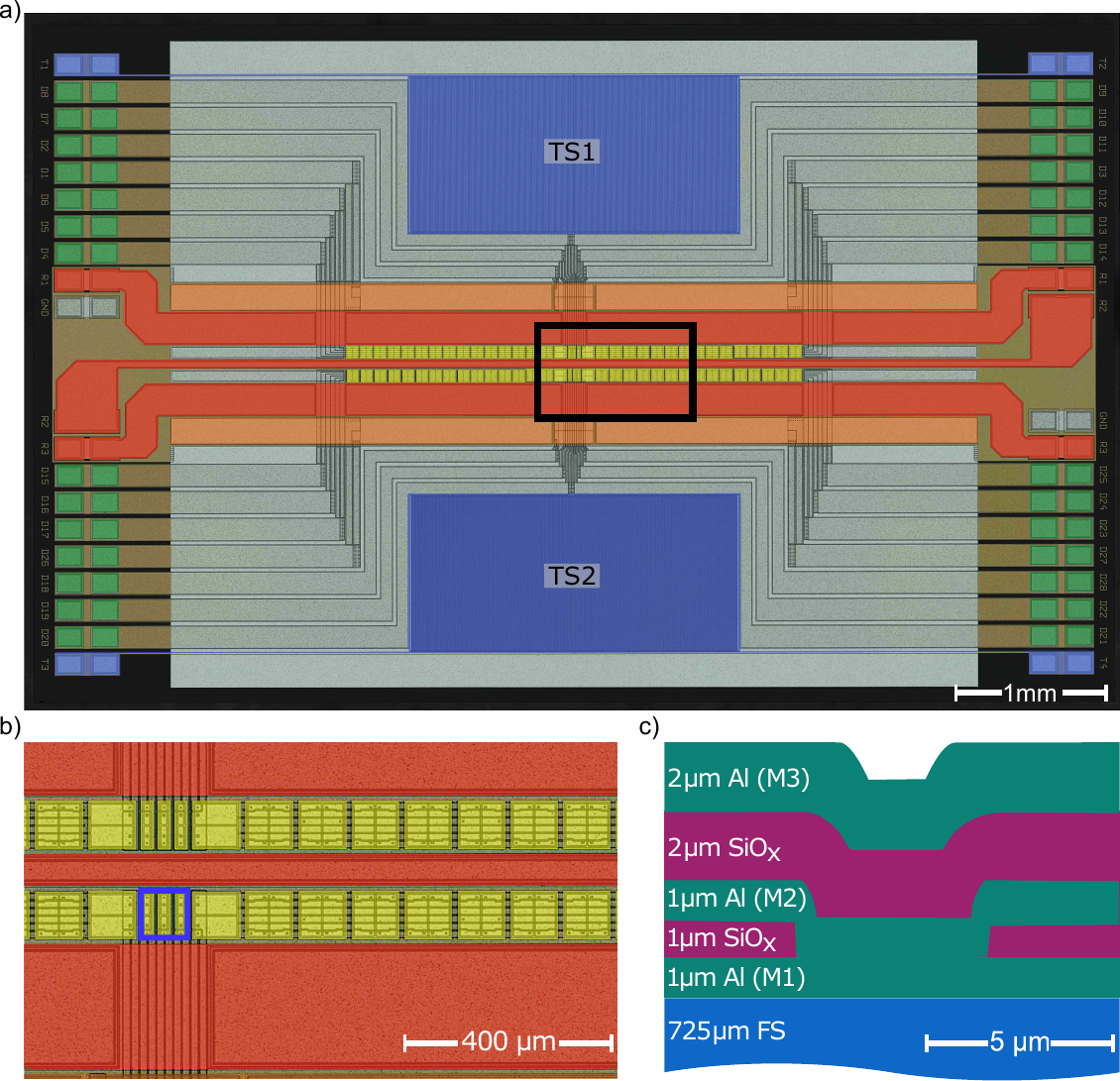}
    \caption{a) Optical microscope picture of the multi-metal layer ion trap used in our experiments, with different colors assigned to highlight different functional areas. The green section represents the divided bond pads for the electrical wafer test, red the RF electrodes, yellow the inner DC electrodes, and orange the outer DC electrodes. The temperature sensors (TS1 and TS2) are colored in blue, located in the first and second aluminum layer. The substrate is visible in black and the grey area is connected to GND. b) Magnification of the black marked section in a) shows three RF rails as well as DC electrodes in between. One DC electrode is marked in blue, indicating the position where ion heating rate measurements (Section \ref{HR}) are conducted. c) Schematic drawing of a cross-section through a connection between the first and second metal layer. The drawing shows the fused silica (FS) substrate (blue) and the ion trap's three aluminum (M1-M3) layers (green), separated by silicon oxide (SiO$_\textrm{x}$, purple).}
    \label{fig:Trap24_overview}
\end{figure}
The trap design, shown in Figure \ref{fig:Trap24_overview}a), resembles an instance of a 2D lattice architecture, more specifically the so-called Quantum Spring Array (QSA) architecture.\cite{valentini2024} In the QSA architecture, ions are distributed in a two-dimensional grid formed by several parallel linear traps placed adjacent to each other. Each of these linear traps has multiple independently controllable trapping sites. Two-dimensional connectivity between ions located in distinct trapping sites is realized by transporting and coupling ions along the directions parallel (axial) and orthogonal (radial) to the RF line.\cite{Tanaka2014, Sterling2014, Bruzewicz2016, Holz2020}
 
The trap layout follows the design proposed by Holz et al.,\cite{Holz2020} which was originally implemented on a silicon substrate. However, in this work, a trap with the same electrode layout is realized on a fused silica substrate. 
It features three parallel RF electrodes (in red), which generate a pseudopotential landscape consisting of two RF minima that resemble an instance of the QSA architecture with two adjacent linear traps. \hl{The outer RF electrodes and the inner one have a width of {\SI{245}{\micro\meter}} and {\SI{64}{\micro\meter}}, respectively, and are separated by {\SI{111}{\micro\meter}}.}
The two pseudopotential minima \hl{created by the RF electrodes} are located at a distance of $\approx$ \SI{125}{\micro\meter} above the trap surface, and separated by $\approx$ \SI{100}{\micro\meter} from each other.
70 DC (direct current) electrodes (in yellow), placed between the RF lines, create 18 trapping sites along the axial direction of the trap, evenly divided between the two RF minima. \hl{The DC electrodes have dimensions of {\SI{95}{\micro\meter}} by {\SI{95}{\micro\meter}} with a gap of {\SI{8}{\micro\meter}} between them}. In the center of the trap, the electrode pitch is reduced from \SI{103}{\micro\meter} to \SI{33}{\micro\meter}, facilitating the generation of anharmonic potentials required to shape the double-well potential in this region. The DC electrodes are co-wired in parallel, with every third electrode being connected \hl{to the same DC supply line}. However, the five central electrodes for each of the two linear traps are routed independently for finer control of the DC potential. Six additional DC electrodes (in orange), located aside the outer RF lines, are utilized for the compensation of stray electric fields. Each DC electrode on the trap is connected to a pair of bonding pads by tracks that run through the lower metal layers, enabling electrical testing of the entire trap chip, as discussed in Section \ref{wafertest}. Two temperature sensors labeled TS1 and TS2 (in blue), are located on the long sides of the trap. The operating principle of the sensors is based on the dependence of the resistance of a long, thin metal line on temperature. More details about the temperature sensors are given in Section \ref{Tsensor}. 

Figure \ref{fig:Trap24_overview}b) shows a detailed view of the part of the trap located in the black rectangle in Figure \ref{fig:Trap24_overview}a). Above the middle of the electrode highlighted in blue, ion heating rate measurements are performed, as discussed in Section~\ref{HR}.
The cross section of the trap is sketched in Figure \ref{fig:Trap24_overview} c). The trap consists of three aluminum metal layers (M1-M3) with oxide (SiO$_\textrm{x}$) deposited in between as insulation and is fabricated on a \SI{725}{\micro\meter} thick fused silica substrate.  The upper layer (M3) contains the trap electrodes, while M1 and M2 are used for routing purposes. 
Vertical connections, called vias, allow electrical signals to be routed across multiple metal layers. In Figure \ref{fig:Trap24_overview} c) the via connects M1 and M2.
Notably, the large bandgap (7.5-\SI{9.6}{e\volt}) and low RF losses ($\tan\delta=0.5\times10^{-4}$) of fused silica\cite{Bussey1964,Tan2003} allow one to remove the GND shield layer, which is typically present as a first layer in silicon-based trap designs to shield the substrate from laser light and RF fields.\cite{Holz2020,Sterk2024,chung2024, Mehta2020, Mehta2014, Niffenegger2020}

When routing electrical signals from the bonding pads to the electrodes, the intersection of RF and DC lines is of particular concern. The capacitive coupling at these intersections can result in an RF pick-up on the DC electrodes, which introduces unwanted variations in the pseudopotential.\cite{Auchter2023,Holz2019} The three-metal-layer ion trap design presented in this work is designed to minimize the RF pick-up on the DC lines. At the intersection points, the RF signal is routed in the top metal layer (M3), the DC signal is routed in the bottom metal layer (M1), and a grounded metal is placed in the middle layer (M2). Finite element simulations indicate that this design reduces the RF-to-DC coupling capacitance from \qty{14}{\femto\farad} without a GND layer to \qty{0.6}{\femto\farad} with a GND layer between RF and DC.

The fabrication of the trap presented here was carried out in the industrial facilities of Infineon Technologies in Villach, Austria. The traps are fabricated on \SI{200}{mm} diameter fused silica wafers with 477 chips per wafer. The fabrication involves 104 process steps, each of which includes loading the wafer into a tool, wafer processing, and wafer unloading.\cite{Franssila2010} In addition to deposition, lithography, and structuring, the fabrication process incorporates several control and cleaning steps. 

The fabrication process is based on Complementary Metal Oxide Semiconductor (CMOS) technology.\cite{Anders2023} The transparency of fused silica poses a challenge during fabrication, as the sensors of most fabrication tools are designed to work with non-transparent substrates, such as silicon.\cite{Haley2024} To enable the handling of the wafers, the fused silica substrate is covered in a \SI{2}{\micro\meter} layer of polycrystalline silicon via chemical vapor deposition\cite{Bryant1977}. Subsequently, the silicon is etched away on the front side of the wafer \hl{using reactive ion etching}\cite{Jansen1996}, leaving only the backside covered with silicon while exposing the bare fused silica on the front side. The \SI{2}{\micro\meter} silicon layer on the backside makes the wafer opaque, thereby enabling its handling by fabrication tools designed for opaque substrates. Once handling is enabled, the ion trap is fabricated directly on the fused silica on the front side of the wafer. After the fabrication process is complete, the remaining \SI{2}{\micro\meter} silicon layer on the backside is removed \hl{by spin etching.}\cite{Yongdae2015}\hl{To prevent any damage to the front side, a resist layer is first applied to protect it. The wafer is then inverted, with the front side facing downwards, and spun. This process allows the etchant to be applied exclusively to the backside. As a result, the silicon layer is removed from the backside without affecting the front side of the wafer,} leaving only the fused silica as the final substrate.

The metal layers are deposited using sputter deposition,\cite{Martin2010} while the silicon oxide (SiO$_\textrm{x}$) layers are deposited through plasma-enhanced chemical vapor deposition.\cite{Martin2010a} The structuring process involves  optical lithography techniques followed by plasma etching.\cite{Donnelly2013} The lithography is a three step process consisting of resist spinning, exposure with \SI{365}{\nano\meter} light,\cite{Yang1991} and resist development. Concerning the exposure, a mask (reticle) is used, which houses seven identical chips and two areas for test structures on a 3x3 grid.\cite{Grobman2001,Milner1994} This reticle is repeatedly exposed over the wafer.\cite{Brink1996}

The thickness of the metal layers determines the minimum distance between adjacent metal structures. The aspect ratio of the etched gap, defined as the ratio between the layer thickness and the gap size, should not exceed 1:1 to ensure reliable fabrication. For the fabrication processes used, experience has shown that at higher aspect ratios, residual metal might remain in the gap after etching, resulting in unintended connections between neighboring metal structures. During fabrication, we do not employ any planarization steps, and therefore the topography adds up with each layer. The accumulation of topography can lead to uneven surfaces, which pose challenges during lithography.\cite{Martinez2006} We therefore choose thin metal and oxide layers to minimize topography.
The first two metal layers are \SI{1}{\micro \meter} thick to allow for small gaps and minimize topography, while the third layer is \SI{2}{\micro \meter} thick to achieve lower resistance due to a larger cross section compared to the other two. Two SiO$_\textrm{x}$ layers, \SI{1}{\micro\meter} and \SI{2}{\micro\meter}, are deposited for isolation between the metal layers. The layers are deposited with a thickness tolerance of ±\SI{10}{\%}. 

The via connecting M1 and M2 (shown in Figure \ref{fig:Trap24_overview}c)), is created by etching a hole in the oxide layer located between the two metal layers. Subsequently, the deposition of M2 fills this hole, establishing a connection between M1 and M2. In-situ argon milling is performed prior to the deposition of the second metal layer. The milling process removes any native oxide that may have formed on top of the first layer, which could compromise electrical connectivity between the metal layers.\cite{Delfino1992}

\subsection{Power dissipation} \label{dissipation_section}
In this section, we explain how and why the usage of fused silica instead of silicon reduces RF-induced power dissipation within the ion trap.
The application of an RF signal to the trap electrodes results in two power dissipation mechanisms: Ohmic losses ($P_{\textrm{Ohmic}}$) and dielectric losses ($P_{\textrm{diel}}$), given by the following expressions
\begin{equation}
    P_{\textrm{Ohmic}}=\frac{V^2_0}{6}C^2R\omega^2, \ \ P_{\textrm{diel}}=\frac{V^2_0}{2}\omega C \tan \delta.  
    \label{dissipation_equations}
\end{equation}
A detailed derivation of these equations is provided in Appendix A.
In Table \ref{Dissipation} we summarize different material parameters and the estimated dissipated power for the three different trap versions, which have the same electrode layout: A partially shielded trap on silicon substrate,\cite{Holz2020} a fully shielded trap on silicon substrate,\cite{Auchter2023} and the trap on fused silica presented in this paper. Both silicon traps are described in more detail in Appendix B.

An insight is that the partially shielded trap exhibits lower power dissipation compared to the fully shielded trap. This is because the fully shielded trap features a larger GND pad (as shown in Appendix B), which increases the RF to GND capacitance, resulting in higher losses, in agreement with equation \ref{dissipation_equations}. Furthermore, in both silicon traps, Ohmic losses $P_{\textrm{Ohmic}}$ dominate over dielectric losses $P_{\textrm{diel}}$ at room temperature, while at cryogenic temperatures, the dielectric losses become the dominant part.
For the fused silica trap, dielectric losses dominate the dissipated power at both room and cryogenic temperatures. More specifically, at room temperature 61\% of the dissipated power is due to dielectric losses. At cryogenic temperatures, dielectric losses account for 98\% of the total power dissipation.
Moreover, the fused silica trap dissipates an order of magnitude less total power compared to the silicon traps at room temperature and a factor of three to five less at cryogenic temperatures. This reduction can be attributed to several factors. 
First, since the fused silica trap does not require a shielding layer between the RF electrodes and the substrate, the RF to GND capacitance is reduced from \SI{28}{\pico\farad} (partially shielded) or \SI{42}{\pico\farad} (fully shielded) to \SI{12}{\pico\farad}. Since $P_{\textrm{Ohmic}}$ scales quadratically and $P_{\textrm{diel}}$ linearly with capacitance, the reduced capacitance influences both dissipation mechanisms. 
Second, the removal of the shield plane allows the RF electrode to extend across all three metal layers. As a result, its thickness increases from \SI{2}{\micro \meter} in the shielded traps to \SI{4}{\micro \meter} in the fused silica trap.  With a larger cross-section, the resistance of the RF electrodes is reduced by a factor of 2, thus reducing $P_{\textrm{Ohmic}}$ by the same factor. 
Third, the utilization of aluminum as opposed to an aluminum-silicon-copper alloy used for the silicon traps reduces the resistivity at \SI{10}{\kelvin} from \SI{2.4e-9}{\ohm\meter}\cite{Holz2020} for our previous traps to \SI{4.3e-10}{\ohm\meter} for our current trap, further reducing $P_{\textrm{Ohmic}}$.
\begin{table*}\centering
\caption{Calculated dissipated power of three trap versions: Partially shielded silicon as described by Holz et al.,\cite{Holz2020} fully shielded silicon as a modification of the first trap with a GND shield covering the entire chip area,\cite{Auchter2023} fused silica without any shielding. All calculations are performed with $\omega\ =\ 2\pi\ \times\ $\SI{22}{MHz}, $V_0\ =\ $\SI{160}{V} and $\tan \delta\ =\ 10^{-3}$. The capacitance of each trap was simulated with finite element simulations, the resistance is derived by the resistivity of the metal and the RF geometry. \hl{The total dissipated power ($P_d$) is calculated by $P_{\textrm{Ohmic}}+P_{\textrm{diel}}$}.}
\begin{tabular}{llllllllll}
    \hline
     Trap & Capacitance (pF) & \multicolumn{2}{c}{Resistance ($\Omega$)}  &\multicolumn{2}{c}{$P_{\textrm{Ohmic}}$ (mW)} & $P_{\textrm{diel}}$ (mW) & \multicolumn{2}{c}{$P_{d}$ (mW)}\\
     Temperature & & \SI{300}{\kelvin} &\SI{10}{\kelvin}&\SI{300}{\kelvin}&\SI{10}{\kelvin}& & \SI{300}{\kelvin} & \SI{10}{\kelvin} \\
     \hline
     Silicon, partially shielded \cite{Holz2020}& 28& 3.0 & 0.31 & 190 & 20 & 50 & 240 & 70\\
     Silicon, fully shielded \cite{Auchter2023} & 42 & 3.0 & 0.31 & 430 &  45 & 74 & 504& 119\\
     Fused silica, no shield & 12 & 1.1 & 0.025 & 13 &  0.3 & 21 & 34 & 21.3 \\
     \hline
     
\end{tabular}
\label{Dissipation}
\end{table*} 

The power dissipation of the presented traps is indirectly estimated via a PT100 temperature sensor placed on the RF resonator housing. Although the resonator housing temperature might also be heated by other sources (like the resonator coil), the observed temperature differences during testing suggest a correlation between trap design and power dissipation.
Under typical application of an RF drive, the base temperature increases from \SI{10(1)}{\kelvin} to \SI{50(1)}{\kelvin} in the partially shielded trap, \SI{20(1)}{\kelvin} in the fully shielded trap, and \SI{11(1)}{\kelvin} in the fused silica trap. The measured temperatures suggest a lower power dissipation in the fully shielded trap compared to the partially shielded trap, contradicting the predictions of the model. We attribute the high temperature measured for the partially shielded trap to the absorption of RF fields in the unshielded region, where the RF field penetrates the silicon substrate. The loss tangent of the low resistivity silicon used for the partially shielded trap is estimated to be $\approx$1.\cite{Yang2006,Krupka2015} The absorption of the RF field in the silicon then leads to heating of the substrate/trap.
The model does not include RF absorption in silicon, and this might explain the observed discrepancy between the experimental results and the model predictions.
With a calculated average power dissipation of \SI{21.3}{\milli\watt} at \SI{10}{\kelvin} and \SI{34}{\milli\watt} at \SI{300}{\kelvin}, the calculations indicate that switching to a fused silica substrate, combined with the removal of the shield layer, represents a valid strategy to reduce power dissipation within ion traps. In fact, our dissipated power is comparable to that of other traps optimized for low power dissipation as shown in the works by Sterk et al. (\SI{28.7}{\milli\watt}),\cite{Sterk2024} and Meinelt et al. (\SI{12.16}{\milli\watt}).\cite{Meinelt2024} Future optimization for power dissipation in ion traps should focus on addressing dielectric losses in the SiO$_\textrm{x}$, as they represent the major contribution to the losses, particularly in fused silica traps at \SI{10}{\kelvin}. There are several approaches to achieve this. One method involves reducing capacitance by increasing the thickness of the SiO$\textrm{x}$ layer or by increasing the distance between RF and other metal structures in the trap design phase. Another potential solution to minimize dielectric losses is the partial removal of the SiO$_\textrm{x}$ layer between the RF and GND layers.\cite{Sterk2024} By removing the SiO$_\textrm{x}$, the dielectric loss tangent in these regions effectively becomes 1, thus reducing the associated dielectric losses.

\section{Electrical wafer test for ion trap reliability} \label{wafertest}
\begin{figure}[t]
    \centering
    \includegraphics[width=\textwidth]{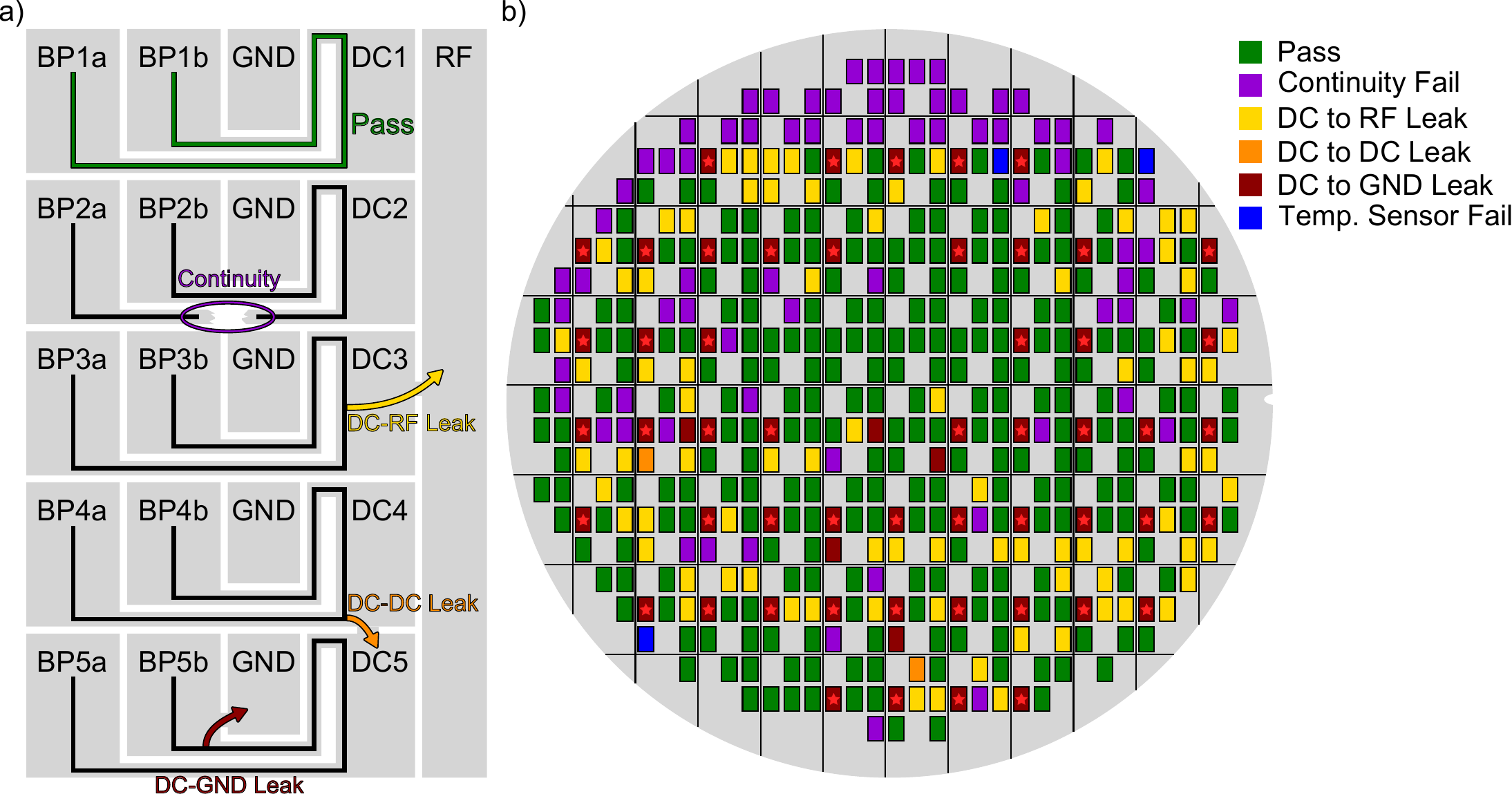}
    \caption{a) Illustration of the loop electrode concept: Bond pad (BP) 1a and 1b are connected through the loop line and DC1. The structural integrity of all parts is verified by measuring the connection from BP1a to BP1b (green). On the loop structures for DC2 to DC5, several types of failures that can be identified through the electrical wafer test are shown: Continuity fail (purple), DC to RF leakage (yellow), DC to DC leakage (orange), DC to GND leakage (dark red). The electrical fail of the temperature sensor is not depicted. b) Electrical wafer test results of a 200 mm wafer with 477 ion trap chips. The black lines indicate the size of a reticle shot. Every reticle shot houses 7 productive chips (squares). Different colors on the wafermap indicate pass (green) or fail (other colors, according to legend) for each trap chip during the electrical wafer test. A red star marks the chips which fail the leakage test and fit into a 3x3 grid, which corresponds to the 3x3 size of one reticle shot. }
    \label{fig:Ringelectrode}
\end{figure}

Fabricating ion traps involves numerous processes, each susceptible to errors. This section introduces an electrical testing protocol to verify the functionality of ion traps after fabrication.
The yield $Y$ represents the fraction of successfully fabricated chips out of the total number of chips on a wafer and depends on the number of critical defects on a wafer. A critical defect is defined as a fault that leads to a malfunction of an ion trap. An example of a critical failure would be an electrical connection between two metal lines that should not be connected (short) or the interruption of an electrode lead (continuity failure). Assuming that all defects are randomly distributed over the wafer, the yield can be described by the Poisson yield model:\cite{Michalka1990}
\begin{equation}
    Y=e^{-N_d/k}
    \label{Poisson}
\end{equation}
where, $N_d$ is the total number of critical defects on a wafer and $k$ is the total number of chips on the wafer. Despite efforts to reach a yield of \SI{100}{\%} in the semiconductor ecosystem,\cite{Donaghy1995,Cunningham1995, Chien2007,Kouta2017,Chien2012,Lee2022} this target is hard to achieve due to contamination, process deviations, substrate defects, film non-uniformity, electrostatic discharge, mask defects, alignment errors, human errors, and more. Therefore, a final validation step is required after all fabrication steps to test the functionality of all chips on a wafer.

Once fabricated, the ion traps can be tested by optical inspection or electrical testing. The optical inspection is limited to the uppermost layer and is thus only viable for single layer ion traps. However, for multilayer ion traps, defects located in the lower metal layers are not visible after fabrication. Therefore, electrical testing is the preferred method for multilayer ion traps.

There exist studies documenting electrical evaluations of ion traps through RF dissipation measurements,\cite{Maunz2019} electrical resistance measurements,\cite{Holz2020} RF breakdown characterization,\cite{Wilson2021} or fault localization,\cite{kehayias2024} yet these do not specifically address the reliability nor assess the functionality of ion traps prior to actual ion trapping, an aspect that is crucial when working with complex ion trap designs. 
Additionally, the results of an electrical wafer test provide extensive statistical information about process deviations and the stability of fabrication processes, allowing one to pinpoint failures, find solutions, and improve ion trap fabrication.

For these reasons, we have developed an ion trap electrical test protocol, with the aim of identifying fabrication errors and drawing conclusions about the fabrication process based on electrical characterization data.
The protocol consists of three types of electrical tests: Continuity tests to verify connectivity, leakage tests to detect unintended electrical connections, and resistance tests to identify irregularities in electrode fabrication. 

The tests utilize 120 probe needles, each having a tip size of \SI{40}{\micro\meter}, allowing one to individually access each bond pad present on the ion trap. \hl{The tests are conducted under ambient conditions, with the tool operating at atmospheric pressure and at a temperature of {\SI{295}{\kelvin}}.}
During the test, needles are placed on each chip to establish electrical contact between the measurement tool and the ion trap.
Direct contact of a probe needle with the metal surface leads to an indent and other deformations of the surface.\cite{Chang2024,Chang2011,Liu2008} To avoid this type of deformation on trap electrodes, which would result in unpredictable alterations of the trap potential, we avoid placing the probe needle and electrode in direct contact.
Instead, each electrode of the ion trap is connected to two bond pads (see Figure \ref{fig:Ringelectrode} a)), forming a conducting loop. In the following, we denote this as a "loop electrode" structure.  
The loop electrode structure enables testing of connectivity between the bonding pads and the electrodes without direct contact of the needles with the electrode surface. This is achieved by measuring the connection between the bond pad BP1a and BP1b, as shown in Figure \ref{fig:Ringelectrode}a). This is what we refer to as the continuity test. A passed continuity test between these two bond pads indicates an electrical connection between BP1 and the electrode DC1, as the bond pads are connected exclusively via the electrode DC1. If an electrode lead is broken (see Figure \ref{fig:Ringelectrode} a) purple) the test fails. 

\begin{table*}\centering
\caption{Summary of electric tests, the forced currents, forced voltages, and the limits for passing the test.}
\label{Limits}
\begin{tabular}{lrrr}
    \hline
    Test & Force & \multicolumn{2}{c}{Limits for pass} \\
    \hline
    Continuity Test & \SI{1}{\milli\ampere} & 0 - \SI{100}{\milli\volt} & 0.8 - \SI{1.3}{\milli\ampere}\\
    Leakage Test (DC) & \SI{50}{\volt} & -100 - \SI{100}{\milli\volt} & 0 - \SI{100}{\nano\ampere}\\
    Leakage Test (RF) & \SI{300}{\volt} & -100 - \SI{100}{\milli\volt} & 0  - \SI{100}{\nano\ampere}\\
    Resistance Test (DC \& RF ) & \SI{5}{\milli\volt} & \multicolumn{2}{c}{0 - \SI{50}{\ohm}} \\
    Resistance Test (TS1) & \SI{5}{\milli\volt} &  \multicolumn{2}{c}{28.9 - \SI{35.7}{\kilo\ohm}}\\
    Resistance Test (TS2) & \SI{5}{\milli\volt} &  \multicolumn{2}{c}{10.3 - \SI{11.3}{\kilo\ohm}}\\
    \hline
\end{tabular}
\end{table*} 

During the continuity test, a current of \SI{1}{\milli\ampere} is applied (forced) through a needle, while voltage and current are measured using a 'sense' needle. The test is passed if the measured voltage falls within the range of 0 to \SI{100}{\milli\volt}  (see Table \ref{Limits}), which corresponds to a maximum loop electrode resistance of \SI{100}{\ohm}. This resistance threshold indicates proper electrode connectivity. As the voltage test is subject to fluctuations in the current source, the current is measured simultaneously with the voltage to monitor these changes. The current test is bound to test limits ranging from \SI{0.8}{\milli\ampere} to \SI{1.3}{\milli\ampere}. If the measured current deviates from this range, the test is stopped and the measurement is flagged as a hardware failure of the measurement device. This test is done on all DC loop electrodes. Furthermore, the same test is applied to the temperature sensor and RF lines.

After the continuity test, the leakage test is performed. Here, unwanted electrical connections between separate electrodes and breakdowns are probed. During this test, a voltage of \SI{50}{\volt} is applied to all DC electrodes and the GND area, except for the electrode under test. The test is passed if the current measured at the tested electrode does not exceed \SI{100}{\nano\ampere} and if the voltage difference between the tested electrode and the measurement ground remains within the range of -\SI{100}{\milli\volt} to +\SI{100}{\milli\volt}. These current limits set an upper threshold for the leakage current through the oxide. The test would fail if the quality of the oxide is poor because of process instabilities or impurities. For RF electrodes, a DC voltage of \SI{300}{\volt} is applied to the RF electrode under examination while sensing on all other electrodes. \hl{This voltage is sufficiently low to prevent breakdown events through air between neighboring electrodes.{\cite{Kemas2018,Peschot2015}}} For the test on the RF electrodes the same test limits as for DC electrode leakage tests apply. Leakage tests are performed twice to identify any shorts that may arise from breakdown events during the first assessment. We did not observe such a breakdown event during testing. 

In the final stage of the testing protocol, the resistance test is performed. In this test, the resistance of all DC loop electrodes, RF electrodes, and temperature sensors are individually measured. The resistance of the components is determined by applying a voltage of \qty{5}{\milli\volt} to the force needle and measuring the resulting current.  
This resistance measurement is used to identify irregularities during fabrication, as deviations in resistance might be caused by over-etched metal structures, metal contamination, or problems during lithography.
The limits for resistance tests are currently set to two standard deviations of the first five tested wafers for the temperature sensor and \SI{50}{\ohm} for every other electrical connection. With further changes in processes (e.g. deposition temperature or purity of metal), stricter resistance test limits may be defined to ensure a more reliable production. \hl{A more comprehensive schematic representation of the wafer test setup is provided in Appendix C.}

Overall, the whole test for one ion trap consists of 480 steps, which are performed in roughly 7.8 seconds. If a single test fails, all measurements for that particular ion trap are stopped and the trap is labeled with a failure code. This label indicates the exact point of failure during the testing process. Although the abortion of the test may introduce a bias on the result, as traps failing e.g. the continuity test would likely fail other tests, the abortion is essential to safeguard the measurement equipment from potential damage.

Figure \ref{fig:Ringelectrode}b) illustrates the results of the wafer test for the batch from which the ion trap, whose characterization is presented in Section \ref{characterization}, was manufactured. The data, encoded in a color scheme, reveal that 29.8\% of traps on the wafer failed the leakage tests, with a noticeable threefold periodicity in the distribution of DC to GND leakage errors (see Figure \ref{fig:Ringelectrode} b) red star). 

The observed error distribution strongly suggests a defect in the lithography mask. Specifically, the periodicity aligns with the layout of the 3x3 chip matrix of the used reticle (see Figure \ref{fig:Ringelectrode}b, black lines). This pattern implies that a localized degradation or defect in the reticle could cause recurring errors at the corresponding position in each reticle shot.\cite{Rider2022}
In addition, the continuity test has a failure rate of 15.5\%, predominantly at the edge of the wafer, indicating that process deviations during one or more fabrication steps are more pronounced at the edges of the wafer.\cite{Tarun2020,Zhang2007} With these results, the fabrication processes will be optimized for the next fabrication run.

The total yield for this wafer is \SI{54}{\%}, corresponding to 258 working chips out of 477. This yield translates to an average of 2.7 defects per process step on the wafer. These results highlight the importance of wafer testing as a final step to validate the functionality of an ion trap prior to its integration into a setup.

\section{Trap characterization} \label{characterization}
\subsection{Temperature Sensor Characterization} \label{Tsensor}
\begin{figure}
    \centering
    \includegraphics[width=\textwidth]{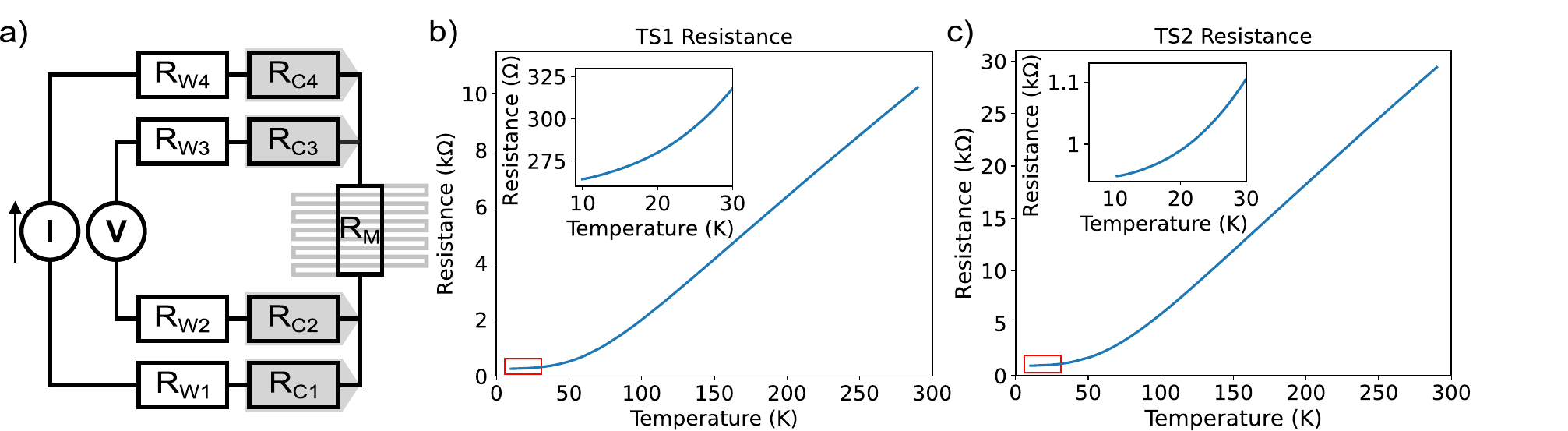}
    \caption{a) Schematic of the temperature sensor test concept. R$_M$ represents the resistance of the meander, while R$_{C1-4}$ describes the contact resistances between needles and pads, and R$_{W1-4}$ refers to the resistance from the wires.  b) and c) show the resistance curve of the two integrated temperature sensors (TS1 and TS2) over the range of 10 to \SI{300}{\kelvin}. The insets of Figures 3b and 3c provide a magnified view of the region between 10 K to 30 K, denoted by the red area.}
    \label{fig:Temperatursensor}
\end{figure}
Monitoring the temperature of an ion trap during its operation is beneficial, as surface noise and thus motional excitation of ions scale with the trap surface temperature.\cite{Bruzewicz2015}
In previous experiments, the trap-temperature was measured through a temperature sensor diode placed on top or near the ion trap \cite{Sage2012,Labaziewicz2008a} or by means of a thermal imaging camera.\cite{Noel2019,Dolezal2015} However, these methods are incompatible with the simultaneous operation of the trap.

To enable temperature tracking during trap operation and precise estimation of the dissipated power on the chip, a temperature sensor is integrated. 
Conceptually, the sensor is based on the dependence of the resistivity of aluminum with temperature. It consists of a long and thin meandering structure. This choice is beneficial since the longer and thinner the meander, the higher the resistance and, ultimately, the higher the sensor's temperature sensitivity ($\Omega K^{-1}$). A higher sensitivity simplifies the measurement process because it amplifies the resistance changes associated with small temperature variations. The sensors are located in the first and second aluminum layers, and are shielded by a GND cover in the third metal layer to prevent any influence of the sensor on the ion.

Two different sensor designs are integrated in the trap, which we refer to as TS1 and TS2 as shown in Figure \ref{fig:Trap24_overview} a). The trace widths of the meanders are \SI{2}{\micro\meter} (TS1) and \SI{1}{\micro\meter} (TS2). 
The sensors have a height of \SI{1}{\micro\meter} and lengths of \SI{70.5}{\centi\meter} (TS1) and \SI{93.5}{\centi\meter} (TS2). Each sensor connects to two bonding pads at each end, allowing one to perform resistance measurements in a four-wire configuration \cite{Smits1958}.

The resistance characterization of the sensors at different temperatures is performed by applying a known current (\SI{5}{\milli\ampere}) to the sensor and simultaneously measuring the voltage, as shown in Figure \ref{fig:Temperatursensor} a). Characterization measurements were taken in a cryogenic test setup equipped with four needles for electrical tests. In this setup, the ion trap is placed on a copper socket, and the temperature is monitored with a temperature sensor mounted directly onto the socket. During the cooling process, the ion trap was cooled at a rate of \qty{1}{\kelvin\per\minute}, while the temperature was measured every 3 seconds. Figures \ref{fig:Temperatursensor} b) and c) display the resistance of the sensors from \SI{10}{\kelvin} to \SI{300}{\kelvin}. These resistance profiles adhere to the conventional metallic resistance model,\cite{Grueneisen1933, Kittel2005, White2019} showing a nearly linear resistance increase between \SI{80}{\kelvin} and \SI{300}{\kelvin}. Below \SI{80}{\kelvin}, the resistance exhibits a lower sensitivity to changes in temperature. However, a monotonic decrease in resistance is observed for both TS1 and TS2 even at low temperatures. Between \SI{10}{\kelvin} and \SI{15}{\kelvin}, the response of the sensor is approximated as linear. In this temperature range the resolution of the sensors is 1.0(5) $\frac{\Omega}{K}$ and 2.5 (5)$\frac{\Omega}{K}$ for TS1 and TS2, respectively. The increased sensitivity of TS2 is consistent with expectations, given its \SI{30}{\%} longer trace length and narrower width. 
The resistance sensitivity measured in both sensors is sufficient to estimate the temperature with an error margin of \SI{0.5}{K}.

On a wafer with 477 chips, the electrical wafer test shows a mean resistance for TS2 (TS1) at room temperature of \SI{32.3}{\kilo\ohm} (\SI{10.8}{\kilo\ohm}) with a standard deviation of \SI{1.6}{\kilo\ohm} (\SI{0.2}{\kilo\ohm}). Assuming that the standard deviation has the same distribution at cryogenic temperatures, the cryogenic characterization measurement yields a standard deviation at \SI{10}{\kelvin} temperatures of \SI{49.4}{\ohm} (\SI{5.3}{\ohm}). Since the expected standard deviation at \SI{10}{\kelvin} exceeds the sensitivity of the sensor per Kelvin, each sensor has to be individually calibrated for a reliable temperature estimation, which is common for most cryogenic temperature sensors.\cite{roths2006}

\subsection{Trap performance characterization} \label{HR}
\begin{figure}
    \centering
    \includegraphics[width=0.9\textwidth]{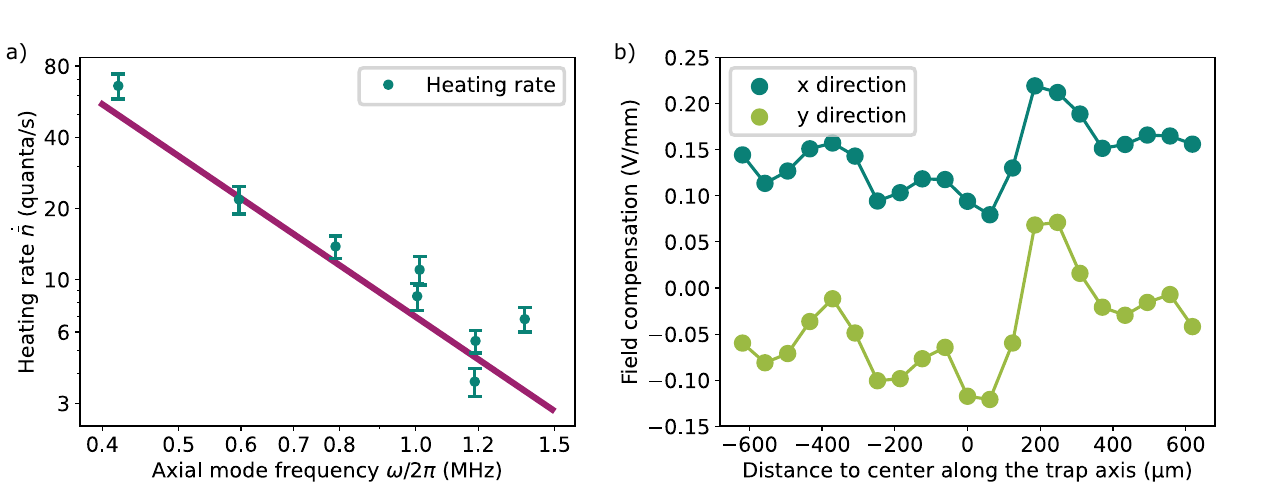}
    \caption{a) Measured heating rates as a function of axial mode frequency at different sites indicated in \ref{fig:Trap24_overview}a). The purple line shows a power law fit ($\dot{\bar{n}} \propto \omega^{-\alpha}$) with $\alpha = 2.0(9)$ to the heating rate data. b) Stray field measurement over a distance of \SI{1.2}{\milli\meter}.}
    \label{fig:HR}
\end{figure}
In this section we evaluate the performance of the ion trap on fused silica, by means of ion measurements with $^{40}$Ca$^{+}$ ions. The trap is characterized by estimating the distribution of electric stray fields along the trap and ion heating rates,\cite{Leibfried2003} i.e. the rate at which the ion motion is excited by sources of electric field noise. The experiments are carried out within a closed-cycle cryostat operated at a base temperature of \SI{10}{\kelvin}.\cite{Niedermayr2015} Details regarding the experimental setup can be found in Holz et al.\cite{Holz2020} The trap is driven with a single RF signal at an RF voltage $V_{RF}=$ \SI{120}{\volt} and a frequency $\Omega_{RF}=$ \SI{17}{\mega\hertz}, resulting in simulated radial frequencies of $2\pi\times$\SI{2.6}{\mega\hertz}. 

Ion heating rates are measured by sideband thermometry on the 4S$_{1/2}$ $\leftrightarrow$ 3D$_{5/2}$ transition.\cite{Leibfried2003} By measuring the ratio of the red and blue sideband transition strength with resolved sideband spectroscopy, we extract the mean motional quantum number. Repeated measurements at different waiting times allow one to calculate the rate at which the ion motion is excited. At each waiting time, the uncertainty arising from quantum projection noise is taken into account.\cite{Bergquist1993} The data collected in Figure \ref{fig:HR} a) show ion heating rates measured above the center electrode marked in blue in Figure \ref{fig:Trap24_overview} b) at various axial frequencies. The error bars represent the fitting uncertainty for each heating rate measurement.
The line in Figure \ref{fig:HR} a) represents a least squares fit to the heating rate data using a power-law function of the form $\dot{\bar{n}}\ \propto \ \omega^{-\alpha}$, with $\alpha$ determined to be 2.0(9). As the DC voltages delivered to the trap electrodes are filtered by first-order RC filters with a cutoff frequency of \qty{5}{\kilo\hertz}, the observed heating rate scaling is consistent with the presence of technical (white) noise, filtered by a first-order low-pass filter.
Additional measurements of the ion heating rates at various trapping sites are provided in Appendix D. These measurements demonstrate a consistent level of heating rates across the entire ion trap. \hl{The reduced motional heating rate compared to the silicon trap may be attributed to multiple factors. The reduced power dissipation in the fused silica trap compared to the silicon trap leads to a reduced surface trap temperature, resulting in lower heating rates.{\cite{Bruzewicz2015}} Assuming that the measured temperature in the setup roughly reflects the temperature of the ion trap, a 10-15 fold improvement in heating rates would be expected. However, up to a 60-fold improvement was observed. A possible explanation for this discrepancy is that the silicon substrate in the trap presented by Holz et al.{\cite{Holz2020}} was not fully shielded in the center trapping regions. Specifically, the first metal layer was partially used for routing, resulting in {\SI{5}{\micro\meter}} wide gaps in the shielding layer. This incomplete shielding can lead to the creation of free electron-hole pairs in the surface, generated by laser light. These free charge carriers may contribute to increased ion heating rates, potentially accounting for the difference between the expected and observed improvements.}

For further characterization of the surface ion trap, stray electric fields are measured in radial x and y directions. Here, y represents the axis perpendicular to the trap surface, while x represents the one parallel to it. 
The electric stray fields are measured across a total distance of \qty{1.2}{\milli\meter} along the trap axis, symmetrically with respect to the trap center, with a resolution of \SI{60}{\micro\meter}. 
The stray fields are estimated by adjusting the DC voltages at each point to minimize both the first-order and second-order micromotion sidebands. To indicate how well the micro motion is compensated, the micro motion parameter $\beta$ was extracted for every measurement.\cite{berkeland1998}  For all measurements $\beta$ stayed below 1.
At each position, the voltage sets required to trap a \calciumforty ion are simulated.\cite{electrode-package} By comparing the simulated voltages with the actual voltages applied to compensate for micro motion, the stray fields are derived.
The results are depicted in Figure \ref{fig:HR} b).
The stray fields are changing along the trap in the range of \SI{0.1}{\volt/\milli\meter} with a peak at \SI{200}{\micro\meter} and are comparable to those observed in our previous traps, as reported in Holz et al..\cite{Holz2020}

After three temperature cycles from room temperature to base temperature, ion transport and trapping at the two central trap sites was no longer viable. However, ion trapping remained possible at other trapping sites. This might be due to the presence of stray charges localized in the trap center, connectivity issues within the trap due to thermal stress, delamination or the experimental setup, which would cause an electrode to disconnect from the DC supply. Despite further investigation (see Appendix E), we cannot conclusively attribute this issue to either the trap itself or the setup. The investigation of this problem, which is beyond the scope of the current work, will be conducted using a new setup with enough DC connections to enable testing of each electrode connection via the loop electrode structures from the outside of the cryostat.

\section{Conclusion}\label{chap:conclucsion}
In this study, we presented the design, industrial-scale fabrication, and characterization of a three metal-layer ion trap on a fused silica substrate, tailored to a 2D lattice architecture. 
We estimated the dissipated RF power within the trap from a circuit model and compared it to ion traps with the same electrode design but fabricated on silicon. We developed a wafer test routine that assesses the electrical functionality of each ion trap chip on a wafer prior to its integration into experimental setups. Moreover, we characterized our fused silica trap by measuring ion heating rates and stray fields in several regions of the trap.

The presented ion trap, developed on a fused silica substrate, is expected to exhibit power losses that are a factor of three smaller than those of silicon-based ones. Our results demonstrate that the use of dielectric substrates and highly conductive metals decreases RF power dissipation, which represents a significant step forward over our previous trap versions. Furthermore, our simulations suggest that to further reduce the dissipated power at cryogenic temperatures, the primary focus should be on improving the dielectrics. One potential solution is to reduce the loss tangent $\tan\delta$ of the inter-metal dielectric by modifying the process parameters or selecting different materials.\cite{Baker-Jarvis1998} Another viable approach is the partial removal of the inter metal oxides while maintaining the mechanical integrity of the metal layers.\cite{Sterk2024} 

Additionally, we developed an electrical wafer test setup capable of identifying discontinuity, short, and leakage errors on the trap. On the wafer where the presented trap was fabricated, a yield of 54\% was achieved. The primary failure modes during fabrication were attributed to potential defects in the mask sets, which will be addressed with a new set of masks, and process non-uniformity, which will be improved by analyzing the acquired data and adjusting process parameters. The implementation of a post-fabrication test represents a noteworthy achievement in the development of reliable large-scale ion traps for quantum computing applications. However, it is important to note that the electrical wafer test is limited as it operates at room temperature, whereas the tested traps are used at cryogenic temperatures. Testing of ion traps designed for cryogenic applications in a cryogenic wafer-probe,\cite{West2022} is left for future work.
Another area of interest for more holistic trap testing is represented by the usage of Kelvin-probe force microscopy to detect irregular surface potentials that could lead to unexpected stray fields on the electrode surface.\cite{berkeland1998} 

We have shown that the built-in trap temperature sensor has a sensitivity of 2.5(5)$\frac{\Omega}{K}$ at \SI{10}{\kelvin}, allowing us to estimate the trap temperature down to a few K with a standard multimeter. Although the sensor was not used during ion trapping, due to an insufficient number of DC connections available within the cryogenic system, we plan to use it in a future setup to monitor trap temperature during operation.

A further improvement over the previous trap versions is represented by the results of the trap characterization. We measured ion heating rates ranging from $\dot{\bar{n}}\approx$ 3 quanta/s at an axial frequency of $\omega = 2\pi$ \qty{1.2}{\mega\hertz} to $\dot{\bar{n}}\approx$ 60~quanta/s at $\omega = 2\pi$ \qty{0.45}{\mega\hertz}. This represents up to a 60-fold improvement over the heating rates measured in silicon-based trap versions.\cite{Holz2020} 

\newpage
\section{Appendix A: Power Dissipation}
This appendix discusses the theoretical aspects of power loss mechanisms occurring in surface ion traps. The application of an RF signal to the trap electrodes results in two power dissipation mechanisms: Ohmic losses ($P_{\textrm{Ohmic}}$) due to the current ($I$) that charges the RF electrode, modeled as a resistor with resistance $R$ in series to a capacitance $C$, and dielectric losses ($P_{\textrm{diel}}$) within the dielectric material. The latter are described by the dielectric loss tangent:\cite{Hughes2011}
\begin{equation}
    \tan\delta=\frac{\epsilon''}{\epsilon'} \, .
\end{equation}
where $\epsilon'$ and $\epsilon''$ are the real and imaginary components of the complex permittivity.
\begin{figure}[b]
    \centering
    \includegraphics[width=0.3\textwidth]{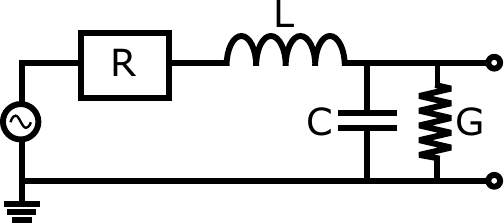}
    \caption{The lumped circuit model with resistance R, capacitance C, inductance L and conductance G.}
    \label{fig:Lumped circuit}
\end{figure}

The total dissipated power $P_d$ within an ion trap is estimated with a lumped circuit model (see Figure \ref{fig:Lumped circuit}), where the dissipated power of the circuit is described by:\cite{Hughes2011}
\begin{equation}
    P_d= P_\textrm{Ohmic}+P_\textrm{diel}=  \frac{V^2_0}{2}Re\left(\frac{1}{Z}\right)
    \label{Dispower_long}
\end{equation}
where $V_0$ is the RF voltage, and the impedance $Z$ is described by the following expression:
\begin{equation}
    Z = R+\frac{G-i\omega C}{G^2+\omega^2C^2}.
    \label{eqn:impedance}
\end{equation}
$\omega$ is the RF drive frequency, $C$ is the RF to GND capacitance and  
\begin{equation}
G=\omega C \tan \delta
\label{eqn:conductance}
\end{equation}
the conductance.

By substituting equation (\ref{eqn:impedance}) and (\ref{eqn:conductance}) into equation (\ref{Dispower_long}), the dissipated power can be expressed as:
\begin{equation}
    P_d=\frac{V^2_0}{2} \frac{R(C^2\omega^2+G^2)+G}{R^2(G^2+C^2\omega^2)+2GR+1}
\end{equation}

For surface ion traps $\omega$ is typically in the range of 10-\SI{100}{\mega\hertz}, the capacitance $C<$ \SI{100}{\pico\farad} and the dielectric loss tangent $\tan\delta <$ 1, it follows that $G\ll$1.\cite{Bruzewicz2019,Sterk2024, Blain2021} Therefore we can approximate $P_d$ with the first order term of its Taylor expansion around G=0:
\begin{equation}
    P_d=\frac{V^2_0}{2}\frac{C^2R\omega^2}{C^2R^2\omega^2+1}+\frac{V^2_0}{2}G\frac{(1-C^2R^2\omega^2)}{(C^2R^2\omega^2+1)^2}+ \mathcal{O}(G^2)
\end{equation}

Given that $CR\omega \ll 1$ (as R is typically $<$\SI{20}{\ohm} \cite{Blain2021}) and using equation (\ref{eqn:conductance}), the dissipated power is approximately:
\begin{equation}
    P_d \approx \frac{V^2_0}{2}C^2R\omega^2+\frac{V^2_0}{2}\omega C\tan\delta
    \label{eqn:dissipation_gleichung}
\end{equation}

where the first term describes the Ohmic losses due to the induced oscillating current, and the second term represents the dielectric losses within the capacitor. 

As the RF electrode works as a capacitor which is charged by the RF current, the distributed nature of the resistance of the RF electrode needs to be taken into account. We consider a model where the RF electrode has a total resistance $R_M=R'L$, which is uniformly distributed along the electrode's length ($L$) with the resistance per unit length $R'$. At the supply point, located at the beginning of the RF electrode, the current is at its maximum ($I_\textrm{0}$). Along the length of the electrode, the current decreases linearly due to capacitive charging. If the voltage drop along the RF electrode is negligible compared to the applied voltage the current is described by 
\begin{equation}
    I(x) = I_{\textrm{0}}\left(1-\frac{x}{L}\right).
    \label{eqn:distributed_current}
\end{equation}
The linear decrease of $I$ leads to non-uniform Ohmic losses along the length of the electrode. Due to the linearly decreasing current along the electrode, the effective resistance $R_{\textrm{eff}}$ is derived by considering the equivalent circuit model. 
The total power dissipation can be obtained by: 
\begin{equation}
    P_{\textrm{Ohmic}} = R'\int_0^L I^2(x) dx=\frac{1}{3}R'LI_{\textrm{0}}^2 = \frac{1}{3}R_\textrm{M}I_{\textrm{0}}^2
\end{equation}
In the lumped circuit model, the power dissipation is given by $P = R_{\textrm{eff}} I_{\textrm{0}}^2 $. Consequently, we conclude that the effective resistance is $R_{\textrm{eff}} = \frac{R}{3} $ when the distributed nature of the resistance is considered.

Substituting $R_{\textrm{eff}}$ for $R$ in equation (\ref{eqn:dissipation_gleichung}) yields the following expressions for the losses:
\begin{equation}
    P_{\textrm{Ohmic}}=\frac{V^2_0}{6}C^2R\omega^2 \ \textrm{ and } \ P_{\textrm{diel}}=\frac{V^2_0}{2}\omega C \tan \delta  
    \label{dissipation_equations_Appendix}
\end{equation}
where P$_{\textrm{Ohmic}}$ only considers Ohmic losses within the RF electrode. Losses in external wiring are not considered here.

\section*{Appendix B: Previous traps on silicon substrate}
\begin{figure}
    \centering
    \includegraphics[width=\textwidth]{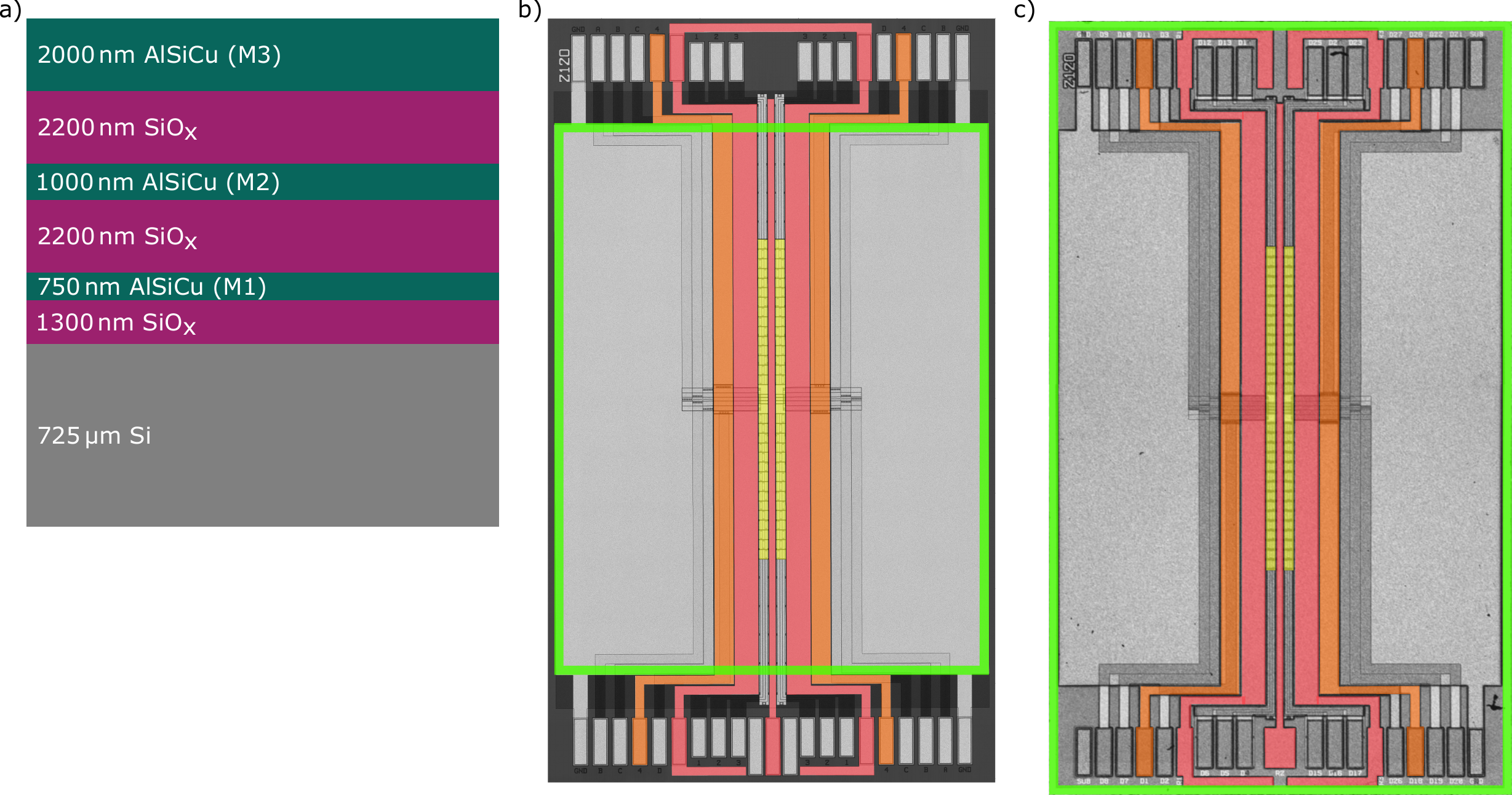}
    \caption{a) Schematic layer stack of the fully shielded and partly shielded ion trap with target layer thicknesses. b) \& c) microscope picture of the partly shielded (b)) and fully shielded (c)) ion trap. The RF lines are colored in red, DC electrodes are yellow, and compensation electrodes are orange. The green marked area indicates the size of the shield layer in the first metal layer.}
    \label{fig:layerstack_appendix}
\end{figure}
In the main text, the fused silica ion trap is compared with a partially shielded\cite{Holz2020} and fully shielded ion trap on silicon.\cite{Auchter2023} A brief overview of these previous traps is provided below, with further details available in the respective references.
Both traps feature three metal layers on a silicon substrate, with a schematic cross-section of the layer stack shown in Figure \ref{fig:layerstack_appendix} a). The traps consist of three oxide layers and three metal layers, with the first oxide layer serving as an isolation layer between the silicon and the first metal layer. The three metal layers (M1-M3) are separated by \SI{2200}{\nano\meter} SiO$_\textrm{x}$. The first metal layer (M1) functions as a ground plane, shielding the substrate from the RF drive and the laser light. The second metal layer (M2) is the routing layer, while the third metal layer (M3) hosts the trap electrodes. \hl{In both silicon traps, the routing of the electrodes located in the center runs through the M1 metal layer. The routing lines are thus isolated from the ground plane via {\SI{5}{\micro\meter}} wide gaps. These gaps are small enough to be neglected for the power dissipation calculation but could lead to the generation of free charge carriers in the silicon substrate.} The trap layout of the partially shielded and fully shielded trap on silicon (Figure \ref{fig:layerstack_appendix} b) and c)) are similar to the design of the ion trap on fused silica (Figure \ref{fabrication_design}) featuring three RF lines (red), two compensation electrodes (orange) and two areas with DC electrodes (yellow), where the trapping sites are located. 

The two traps on silicon differ in two key aspects. First, the fully shielded trap features a ground plane that extends across the entire trap (green), whereas the partially shielded trap lacks a GND layer beneath the bond pad area. Second, the outer RF electrodes' design (red) differs, with the fully shielded trap having two separate RF lines that require individual connections, and the partially shielded trap featuring a single RF line that runs around the trap and can be powered with one connection.

\section{Appendix C: Electrical wafer test schematics}

\begin{figure}
    \centering
    \includegraphics[width=0.9\textwidth]{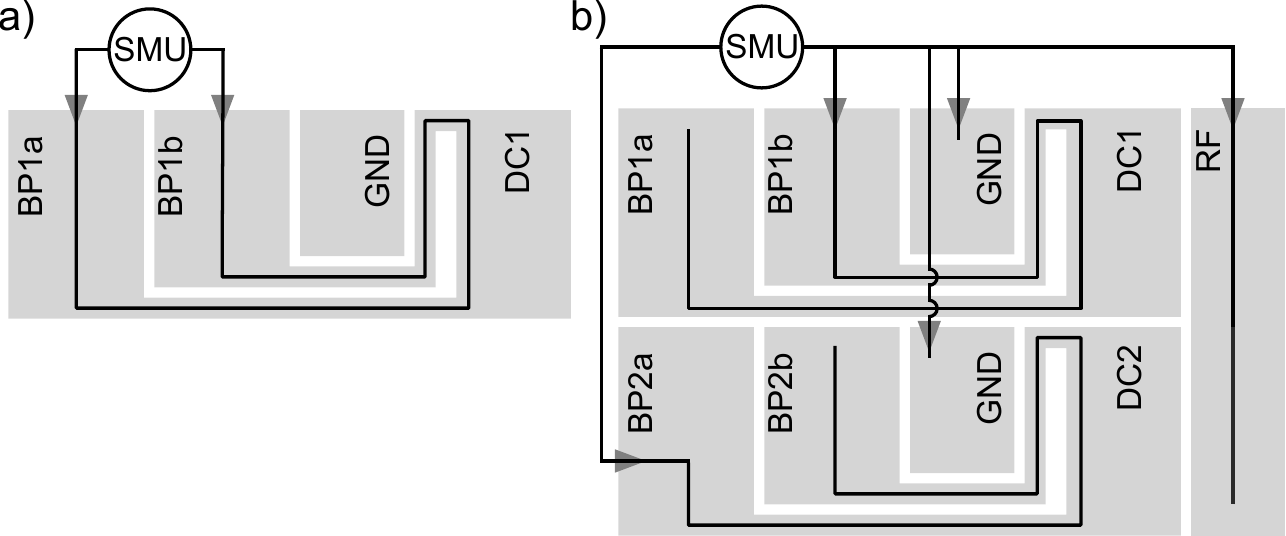}
    \caption{\hl{a) Schematic illustration of the wafer test. During testing, each pad (grey) is contacted by a needle (dark gray), which is connected to one or more source measure units (SMUs). For simplicity, only a single SMU is depicted. a) Continuity test for DC1: the electrical connection between the left and right sides of the SMU is verified. b) Breakdown test for DC2: voltage and current are measured between the left and right terminals of the SMU while a voltage of either {\SI{50}{\volt}} or {\SI{300}{\volt}} is applied to the right side.}}
    \label{fig:electrical_wafertest_explained}
\end{figure}
\hl{This appendix provides a more detailed explanation of the electrical wafer test setup. Each pad on the ion trap chip is contacted via a probe needle. These needles are connected to multiple voltage sources and SMUs through a bus system, enabling fast and flexible measurement configurations.
Figure~{\ref{fig:electrical_wafertest_explained}} shows the two main measurement configurations. In a), a continuity test for the DC1 loop electrode is illustrated. The two bond pads connected to DC1 are probed, and a current of {\SI{1}{\milli\ampere}} is forced through the line while voltage and current are measured to verify the connection. The same setup is used for resistance tests.
Figure~{\ref{fig:electrical_wafertest_explained}} b) shows a schematic of a breakdown or leakage test. In this test, all electrodes except the one under test are biased to {\SI{50}{\volt}} (for DC tests) or {\SI{300}{\volt}} (for RF tests), while the tested pad is held at ground. The SMU then measures the resulting leakage current and voltage. In the example shown, the DC2 pad is tested while all other electrodes (GND, RF, DC1) are ramped to {\SI{50}{\volt}}.}

\section*{Appendix D: Heating rate measurements at different sites}
\begin{figure}
    \centering
    \includegraphics[width=0.7\textwidth]{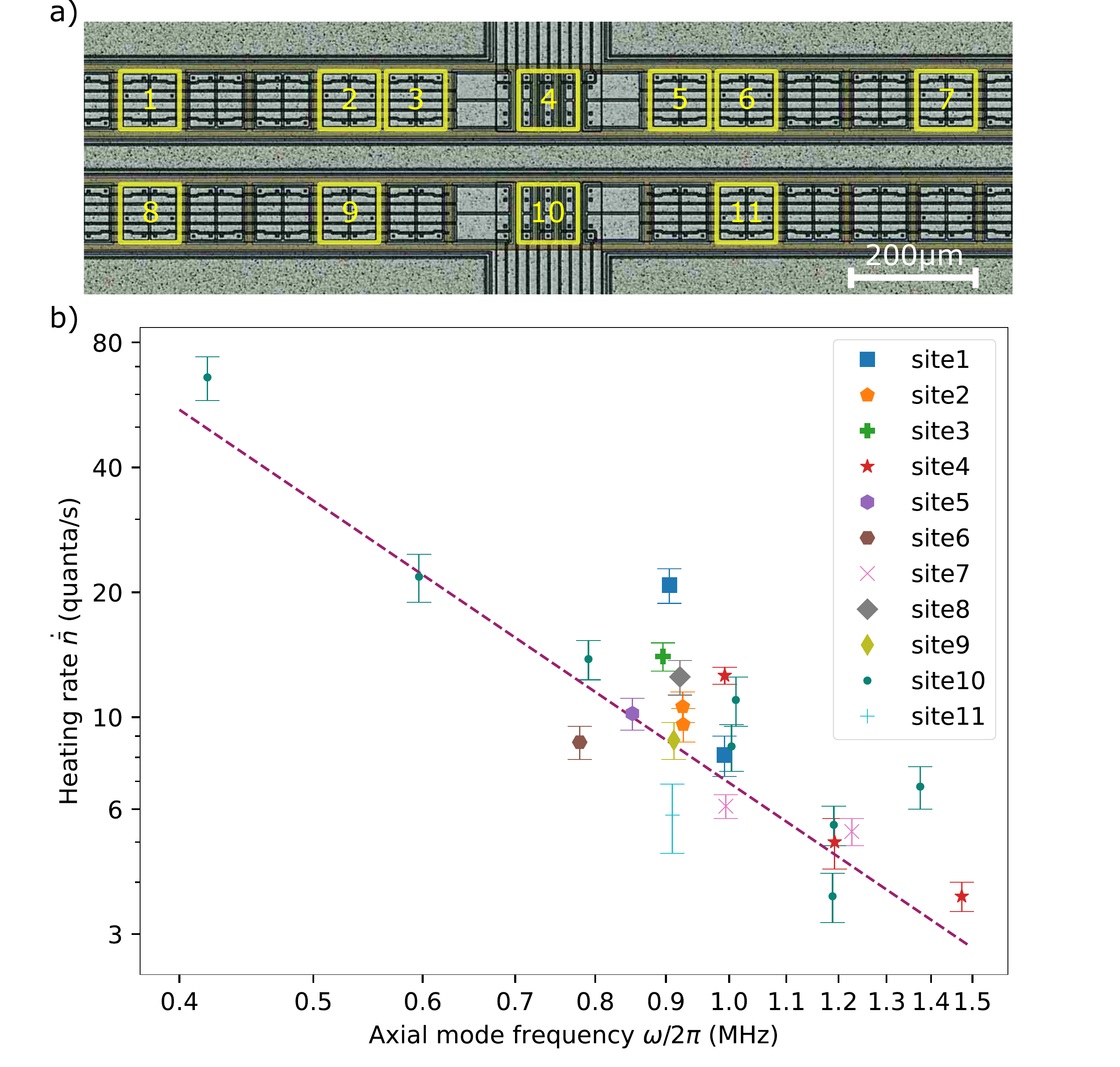}
    \caption{a) Zoom in on the center of the ion trap. At site 1-11 heating rate measurements were taken in b). b) Measured heating rates as a function of axial mode frequency at different sites indicated in a). The purple dashed line shows a power law fit ($\dot{\bar{n}} = \omega^{-\alpha}$) with $\alpha = 2.0(9)$ to the data of site 10 (also shown in fig \ref{fig:HR}).}
    \label{fig:HR_appendix}
\end{figure}
In this Appendix we present ion heating rate measurements performed at 10 additional trapping locations (see Figure \ref{fig:HR_appendix} a) ). The results are summarized in Figure \ref{fig:HR_appendix} b). The purple dashed line shows a linear fit to the heating rate data measured in site 10 as shown in Figure \ref{fig:HR}. Overall, the heating rate data from all sites agree with each other, indicating the uniformity of the heating rates throughout the trap. Table 3 provides all the heating rate data for the fused silica trap. For comparison, the heating rate data of the partially shielded ion trap on silicon taken at the same sites and roughly at the same frequency are also inserted in the table.\cite{Holz2020} In specific regions, such as site 10 or site 4, an improvement of up to 70 times in heating rates is observed.

\begin{table}
    \centering
    \caption{Comparison of ion heating rate data presented in Holz et al. \cite{Holz2020} with the ones measured in this work, for different trapping sites and axial frequencies.}
    \begin{tabular}{lcccc}
    \hline
        site & $\omega$ (FS) & $\dot{\bar{n}}$(FS)& $\omega$ (Si) \cite{Holz2020}& $\dot{\bar{n}}$(Si)\cite{Holz2020}\\
        \hline
         1  & \SI{0.93}{\mega\hertz} & 10.6 (9) &  &\\
         1  & \SI{0.93}{\mega\hertz} & 9.6 (9) &  &\\
         2  & \SI{0.91}{\mega\hertz} & 20.8 (20) &  &\\
         2  & \SI{0.99}{\mega\hertz} & 8.1 (9) & & \\
         3  & \SI{0.90}{\mega\hertz} & 14 (11) & & \\
         4  & \SI{0.99}{\mega\hertz} & 12.6 (6) & & \\
         4  & \SI{1.19}{\mega\hertz} & 5 (7) & & \\
         4  & \SI{1.47}{\mega\hertz} & 3.7 (3) & 1.45 MHz & 288 (35) \\
         5  & \SI{0.85}{\mega\hertz} & 10.2 (9) & & \\
         6  & \SI{0.78}{\mega\hertz} & 8.7 (8) & & \\
         7  & \SI{0.99}{\mega\hertz} & 6.1 (4) & & \\
         7  & \SI{1.23}{\mega\hertz} & 5.3 (4) & 1.24 MHz & 131 (13) \\
         8  & \SI{0.92}{\mega\hertz} & 12.5 (12) & & \\
         9  & \SI{0.91}{\mega\hertz} & 8.8 (9) & & \\
         10 & \SI{0.42}{\mega\hertz} & 65.9 (8) & & \\
         10 & \SI{0.60}{\mega\hertz} & 21.8 (29) & & \\
         10 & \SI{0.79}{\mega\hertz} & 13.8 (15) & & \\
         10 & \SI{1.00}{\mega\hertz} & 8.5 (11) & & \\
         10 & \SI{1.01}{\mega\hertz} & 11 (15) & & \\
         10 & \SI{1.19}{\mega\hertz} & 3.7 (12) & & \\
         10 & \SI{1.19}{\mega\hertz} & 5.5 (6) & & \\
         10 & \SI{1.38}{\mega\hertz} & 6.8 (8) & 1.48 MHz & 472 (50) \\
         11 & \SI{0.91}{\mega\hertz} & 5.8 (11) & & \\
          \hline
    \end{tabular}
    \label{tab:my_label}
\end{table} 

\newpage\section*{Appendix E: Cryogenic fault characterization}
\begin{figure}
    \centering
    \includegraphics[width=0.9\textwidth]{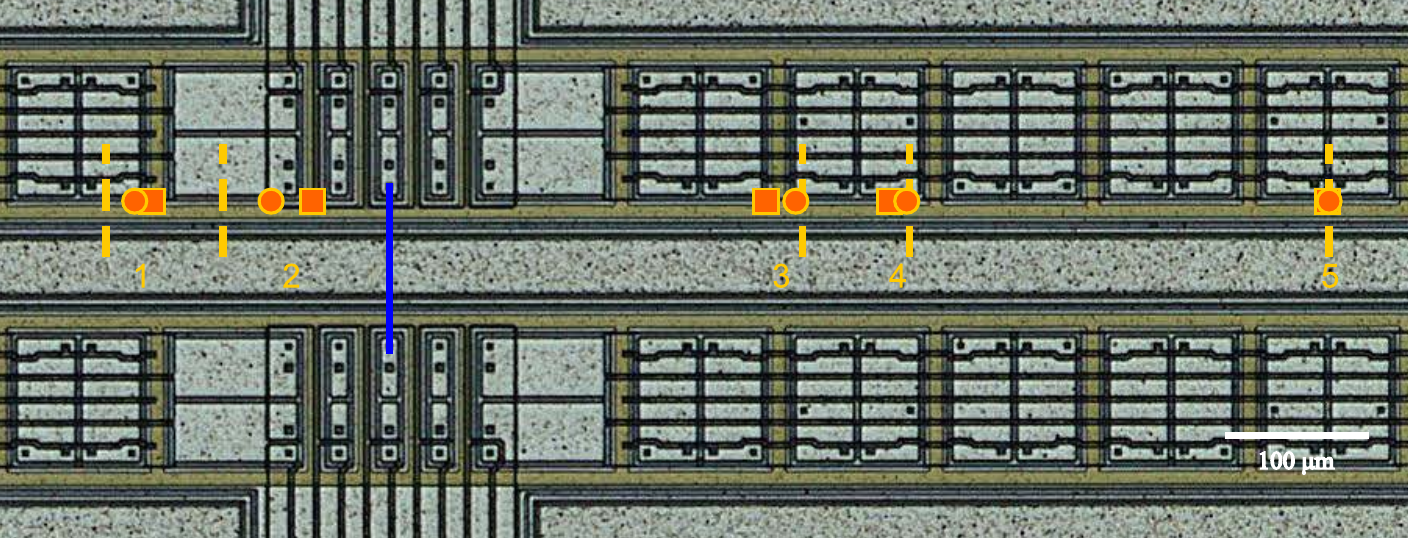}
    \caption{Ion position measurements as function of axial confinement. The measurements are taken for an axial confinement of 1 MHz (squares), and a confinement of $\sqrt{2}$ x 1 MHz (circles), at 5 different trapping positions. The dashed lines represent the expected trapping locations as determined by trap simulations. The blue line indicates the trap center.}
    \label{fig:faultyelectrode}
\end{figure}
In this Appendix we investigate possible reasons behind the degradation of the trap performance related to cryo-thermal cycling. 
As reported in Section \ref{characterization}, after three cooling cycles, trapping was no longer possible at sites 4 and 10 (see Figure \ref{fig:HR_appendix}). However, in their proximity, trapping was possible, but the ions were observed to displace along the trap axis while changing the axial trap frequency. 

This behavior suggests that an additional electric field, not existing prior to the third cryo-thermal cycle, is present along the trap axis. We can identify two main explanations for this effect: Either one or more trap electrodes are disconnected from the DC supply, or the ion displacement comes from stray fields on the trap.

These options are tested taking ion position measurements at different locations along the trap axis, for voltage sets that should cause two different axial frequencies. The measurements are then compared with the ion positions extracted from trap simulations based on the gapless plane approximation.\cite{Gapless2008}  For each axial position, the first voltage set results in a simulated axial frequency of \qty{1}{\mega\hertz}, while the second, equal to the first multiplied by a factor of two, should increase the axial frequency of the ion to $\sqrt{2}$ x \qty{1}{\mega\hertz}, without changing its equilibrium position. 
The measurements performed allow one to localize the region where a floating electrode or stray charges are localized. To distinguish between the two cases, one would have to measure the axial frequency at each ion position along the trap axis, as the voltage applied at the ion position changes if one or more electrodes are disconnected from the supply compared to the case where all the electrodes are connected, and the axial frequency would vary in the two cases.  As the axial frequency was not measured at each ion position, our investigation can clarify the zone in which either a floating electrode or stray charges are present, but it cannot distinguish between the two scenarios.

Figure \ref{fig:faultyelectrode}  shows both the measured and expected ion positions on the trap for the two different axial confinement sets. Although the simulations predict that the position of the ion does not vary with a change in the axial frequency, the ion effectively moves towards the expected trapping location, as the axial confinement is increased. Moreover, in all five different locations, the ion moves outward with respect to the trap center.

From trap simulation, we exclude a short to ground as a possible electrode failure, since a shorted electrode would keep the ion at a constant position if all the voltages are scaled by a constant factor. 

A floating electrode is modeled keeping its voltage constant when the voltages applied to the other electrodes are scaled to change the axial trapping frequency. In order to model stray fields on the trap, we assign a specific voltage to the regions on the trap surface where there is exposed dielectric, i.e. the electrode gaps, and extract the field at the ion position. From trap simulations, we conclude that the observed ion displacement is compatible with the presence of a floating electrode or stray charges on the trap. As the residual field produced by a floating electrode or stray charges is considered constant in time, the relative amplitude of such a field at the ion position will be less the more the voltages of the other electrodes are increased. The position of the total potential minimum thus changes, as observed in the experiment. By symmetry, we can identify the trap center as the area where stray charges or floating electrodes are located. In fact, only a field generated in that location has opposite directions above and below the center of the trap and can exert a force that pushes ions away from the center region. This scenario is compatible with both a floating electrode in the center, which possesses a negative charge, and with negative stray charges located in the trap center. 

As the measurements performed up to now do not allow us to distinguish between the two cases, we conducted three analysis methods outside of our experimental setup to determine whether one of the trap electrodes was floating: Confocal Scanning Acoustic Microscopy (CSAM),\cite{Yu2020} multiple cross sections using Focused Ion Beam (FIB),\cite{Tohru1996} and electrical loop electrode checks at both room temperature and cryogenic temperatures.
During the CSAM analysis, the reflection of acoustic waves at media interfaces and density irregularities are utilized to identify layers that have detached from adjacent layers (delamination) within the ion trap.\cite{Zuliang1995}  Thermal stress could lead to a delamination in the area of the via structures, causing disconnected electrodes. Six ion traps from the same wafer as the one used in this study were examined. We cooled the samples to \SI{77}{K} by submerging them in liquid nitrogen. This cooling procedure, which is repeated 10 times, induces thermal stress in the sample, as it exposes it to a steep temperature gradient.  The CSAM analysis is capable of detecting delaminated areas larger than \SI{25}{\micro\meter} x \SI{25}{\micro\meter}. Therefore the analysis was focused on the RF electrodes, where the metal is connected through all three metal layers. None of the samples showed delamination in the CSAM analysis.
As CSAM can not detect potential delaminations of via structures, as their cross-section is \SI{5}{\micro\meter} x \SI{5}{\micro\meter}, FIB cross-sections were taken at multiple VIA structures to detect any delamination. These cross sections showed an intact metal-to-metal connection in the via and did not show any indications of delamination. 

As a final test, the trap, which displayed trapping issues in the center region, was removed from the setup and examined in an electrical cryogenic test setup. The setup consists of a cryostat equipped with four needles to test electrical connections. The loop electrodes described in Section \ref{wafertest} were used for continuity measurements. All electrodes demonstrated connections at both room and cryogenic temperatures. Additionally, other traps from the same wafer were subjected to multiple cryo cycles in the cryo-prober, and the loop electrodes were checked. All loop electrodes showed resistance $<$ \SI{10}{\ohm} for cryogenic measurements and $<$ \SI{50}{\ohm} for room temperature measurements.

The experimental setup used to trap ions consists of a cryogenic chamber equipped with twenty BNC feedthroughs. The cryostat was heated to room temperature and opened for inspection, to verify the integrity of the electrical connections. The continuity of connections from the outside of the cryostat to the ion trap was tested using a multimeter with two measurement probes. It is not feasible to measure these connections at cryogenic temperatures because the cryostat must remain sealed during the cooling process.  A loop scheme (similar to the loop electrode described in chapter \ref{wafertest}) would be needed to verify the electrical connection from the outside of the cryostat to the ion trap.
However, the current configuration of the printed circuit board (PCB) housing the ion trap is not suited for such loop measurements. The PCB design lacks loop structures, which would allow testing while the cryostat is closed and cooled. Furthermore, the restricted number of electrical connections available inside and outside of the cryostat limits the testing capabilities. 

Despite extensive efforts to identify the issue that prevented trapping in the trap center after multiple cool-down cycles, the underlying cause can not be specifically tied to the setup or the trap. A planned future experimental setup will incorporate 100 DC connections into the cryostat, enabling cryogenic testing of all the loop electrode structures present in the trap during trap operation.

\vspace{1cm}
\noindent \textbf{Acknowledgements} \par 
\noindent We gratefully acknowledge support by the ECSEL JU (which is supported by European Union’s Horizon 2020 and participating countries) under Grant Agreement Number 876659 (iRel), the Austrian Research Promotion Agency (FFG) under the project “OptoQuant” (37798980), 
the European support “Important Projects of Common European Interest” (IPCEI) on Microelectronics,
the European Union’s Horizon Europe research and innovation program under Grant Agreement Number 101114305 (“MILLENION-SGA1” EU Project), 
the European Union’s Horizon Europe research and innovation program under Grant Agreement Number 101046968 (BRISQ), 
the Austrian Science Fund (FWF Grant-DOI 10.55776/F71)(SFB BeyondC), 
and the Austrian Research Promotion Agency under Contracts Numbers 896213 (ITAQC) and 914032 (ScaleQudits).

\vspace{0.25cm}
\noindent We thank Yves Colombe for fruitful discussions during the chip fabrication and during the preparation of the manuscript.

\noindent We thank Nina Megier for help with simulations.

\medskip
\bibliographystyle{MSP}
\bibliography{refs}

\end{document}